\documentclass[
reprint,
amsmath,amssymb,
prx,
longbibliography,
floatfix,
citeautoscript
]{revtex4-2}
\usepackage{amsmath}
\usepackage{braket}
\usepackage{graphicx}
\usepackage{dcolumn}
\usepackage{bm}
\usepackage{tabularx}
\usepackage{makecell}
\usepackage{xcolor}
\usepackage{xr-hyper}
\usepackage{hyperref}

\begin{document}

\preprint{APS/123-QED}

\title{Non-equilibrium Dynamics of Two-level Systems directly after Cryogenic Alternating Bias}

\author{V. Iaia$^1$}
\email{iaia1@llnl.gov}
\author{E. S. Joseph$^2$}
\author{S. Im$^3$$^{,4}$}
\author{N. Hagopian$^3$}
\author{S. O'Kelley$^1$}
\author{C. Kim$^1$}
\author{N. Materise$^1$}
\thanks{Present address: QuantWare BV, Elektronicaweg 10, 2628, XG, Delft, Netherlands}
\author{S. Patra$^1$}
\author{V. Lordi$^1$}
\author{M. A. Eriksson$^2$}
\author{P. M. Voyles$^3$}
\author{K. G. Ray$^1$}
\author{Y. J. Rosen$^1$}
\email{rosen10@llnl.gov}
 
\affiliation{$^1$Lawrence Livermore National Laboratory, Livermore, CA 94550}
\affiliation{$^2$Department of Physics, University of Wisconsin-Madison,}
\affiliation{$^3$Department of Materials Science and Engineering, University of Wisconsin-Madison, Madison, WI 53706}
\affiliation{$^4$Department of Materials Design and Innovation, The State University of New York, Buffalo, NY 14220}

\date{\today}

\begin{abstract}
    Two-level systems (TLSs) are tunneling states commonly found in amorphous materials that electrically couple to qubits, 
    resonators, and vibrational modes in materials, leading to energy loss in those systems. Recent studies suggest 
    that applying a large alternating electric field changes the oxide structure, potentially improving 
    the performance of qubits and resonators. In this study, we probe the effect of alternating bias at cryogenic temperatures on TLS dynamics within amorphous oxide parallel-plate capacitors operating in the strongly coupled regime. We bias the TLSs in the capacitors 
    using an electric field. This allows us to spectroscopically image TLSs and extract their densities and dipole moments. 
    When an in-situ alternating bias is applied, the steady-state spectra from the standard TLS model disappear. Post-alternating bias 
    TLS spectroscopy reveals transient behavior, in which the TLS frequency fluctuates on the order of minutes. Thermal cycling 
    above 10 K reverses these effects, restoring the TLS spectrum to its original state, indicating a reversible mechanism. 
    Importantly, the intrinsic loss tangent of the LC oscillator remains unchanged before 
    and after the application of the alternating bias. We propose that the disappearance of the steady-state spectrum are caused by non-equilibrium energy build up from strain in the oxide 
    film introduced by the pulsed voltage bias sequence. Understanding this non-equilibrium energy could inform future 
    models of time-dependent TLS dynamics.
\end{abstract}\label{abstract}

\maketitle


\section{Introduction}\label{introduction}

Two-level systems (TLSs) are tunneling states in amorphous materials that contribute to energy loss and 
decoherence in superconducting and low-noise rf devices \cite{muller2019,yu2023}. These effects are particularly 
detrimental to superconducting quantum bits (qubits), for which TLSs have been identified as a major source of 
decoherence. In 2005, Martinis et al. \cite{martinis2005} discovered this connection, prompting extensive research 
from the qubit community into their origins and mitigation \cite{muller2019}. Subsequent studies have shown that TLS-qubit 
interactions can persist for hours \cite{klimov2018} and contribute to telegraphic noise and 1/f noise 
\cite{schlor2019correlating, muller2015interacting}. Beyond qubits, TLSs are a significant source of phase 
noise in microwave kinetic inductors used for astronomical photon detection 
\cite{gao2007noise,gao2008physics,mazin2009microwave,noroozian2009two}, and they have been linked to 
charge noise in quantum dots \cite{connors2019low}.

The origins of TLSs have been a subject of investigation for over 50 years \cite{yu2023}. Proposed origins 
include atomic dislocations \cite{dubois2015atomic,dubois2013}, dangling hydrogen bonds \cite{holder2013bulk,gordon2014}, 
and surface or interface states \cite{faoro2012}. TLSs have also been speculated to occur in Josephson junctions 
\cite{martinis2005}, a critical component of superconducting qubits \cite{oliver2013materials}. These junctions feature 
oxide barriers sandwiched between superconducting lines where high densities of TLSs are concentrated and can directly affect 
qubit operation. Additional studies have identified TLSs in substrates, surface oxides \cite{barends2013coherent}, and native 
oxides \cite{deng2013analysis}. Despite extensive research, the precise origins of TLSs remain unclear.

Recent studies have probed the effects of alternating voltage bias on device performance. Several studies have shown that they can manipulate the population \cite{burin2013universal,matityahu2017rabi} and the frequencies \cite{dane2025performance,chen2025scalable} of the TLS bath. Further studies have shown that applying a large alternating voltage bias changes the oxide barrier of Josephson junctions for superconducting qubits, 
offering a potential approach to improving device performance \cite{pappas2024alternating,wang2024precision}. This alternating bias technique has 
been shown to accurately tune qubit frequencies by altering junction resistance. This change in resistance may be due to a non-reversible 
process in the junction caused by migration and diffusion of Al$^{3+}$ and O$^{2-}$ ions within the barrier during the alternating processing. To amplify the effects of this biasing technique, 
previous experiments utilized a heated probing stage at temperatures up to 353 K \cite{kennedy2025tuning,pappas2024alternating,wang2024precision}.

\begin{figure}[h!]
    \centering
    \includegraphics[width=3.3in]{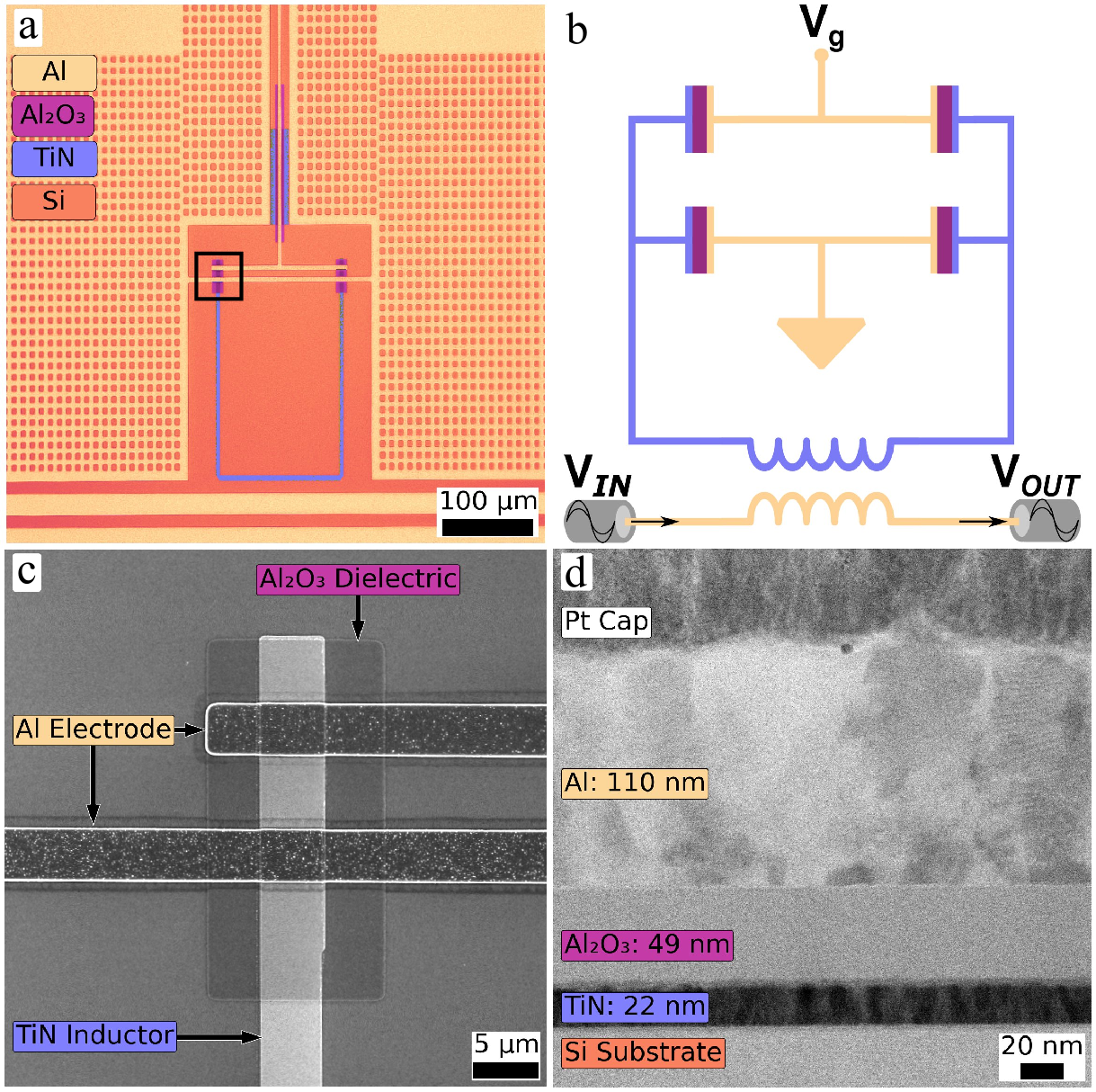}
    \caption{(a) False-colored optical microscope image of the LC oscillator consisting of a Si substrate (orange), 
    TiN (blue), Al$_2$O$_3$ (purple), and Al (yellow) layers. A voltage bias is applied along the thin, vertical Al 
    channel. The TiN inductor is connected to four capacitive elements. The black box depicts two capacitors connected in a bias bridge. When biased, the capacitors have a constant dc 
    field, E$_g$=$V_g$/2$d_o$, where $d_o$ = 49 nm is the dielectric thickness. (b) A circuit diagram for the 
    lumped-element resonator. The color coding depicts the same galvanic connections as in (a). (c) An SEM image of the two capacitors highlighted in (a), 
    acquired on a different oscillator of the same design. (d) A bright field-STEM image of the trilayer 
    TiN-$\rm{Al}_{2}\rm{O}_{3}$-Al capacitor cross-section, capped with a Pt protective layer.}
    \label{fig:Device_image}
\end{figure}

This study investigates the impact of in-situ Cryogenic Alternating Bias Stimulation (CABS) on TLS dynamics in an amorphous aluminum oxide ($\alpha$-$\rm{Al}{_2} \rm{O}{_3}$) film.
We use an rf lumped-element LC oscillator in the strong coupling regime \cite{sarabi2015} to probe TLS dynamics.
Similar platforms have been used to measure TLS dipole moments by adjusting the TLS energy within a large oxide barrier \cite{sarabi2016,hung2022}.
By applying an electric bias across the capacitor, we change the energy of TLSs within the bandwidth of our oscillator. The TLS-oscillator interaction is observed as an avoided-level crossing
described by the standard tunneling model \cite{anderson1972,phillips1972}. Using this model, we extract the electric dipole moment ($\vec{p}$) of the strongly coupled TLSs. 
After we apply CABS, TLSs no longer exhibit the characteristic steady-state spectral signatures they did upon initial cooldown \cite{sarabi2016,hung2022}. Instead, analysis shows that the TLS-induced 
avoided crossings are now irregularly distributed and change in time across the measurement bandwidth. Additionally, loss tangent experiments after each 
voltage and thermal treatment indicate that low-power loss does not change. Furthermore, after cycling to 10 K the TLS spectral signatures return to the pre-CABS state. 

The remainder of this work will be organized as follows: In Section \ref{Methods}, we describe the design and operation of the experimental circuit, highlighting its sensitivity to TLSs 
embedded in trilayer capacitors. A review of our TLS spectroscopy and voltage biasing procedures is provided. In Section \ref{Results}, we present TLS spectroscopy during 
different in-situ CABS and thermal cycles. From these spectra, we extract dipole moments and densities using the standard tunneling model. Loss tangent measurements after 
each voltage and thermal treatment are compared to understand changes in low-power loss and critical photon number. In Section \ref{Discussion}, we discuss the implications of 
CABS on TLS dynamics and loss tangent measurements. Finally, we conclude by summarizing the key findings of TLS densities post-CABS and loss tangent measurements in 
Section \ref{Conclusion}.

\section{Methods}\label{Methods}

\subsection{Fabrication and Device Design}\label{Device Specifics}

We use a lumped-element LC oscillator formed by a parallel-plate bias bridge capacitor \cite{sarabi2016,hung2022,khalil2014landau} and 
an inductive element coupled to a feedline, as shown in Fig. \ref{fig:Device_image}. The device was fabricated on high-resistivity 
silicon using three lithographic steps. The bias bridge capacitor structure is a trilayer consisting of a 22 nm TiN film, a 49 nm 
amorphous aluminum oxide ($\alpha$-$\rm{Al}{_2} \rm{O}{_3}$), and a $\sim$110 nm Al capping layer [see cross-section image in Fig. \ref{fig:Device_image}(d)]. 
The TiN layer were deposited using plasma-enhanced atomic layer deposition (ALD). The oxide layer were deposited using thermal ALD, while the Al layer was added via electron beam evaporation. 
It should be noted that the ALD-grown oxide barrier in this study differs from the commonly employed oxide growth method for Josephson junction-based qubit fabrication, which is a diffusion-limited oxidation of an aluminum layer \cite{oliver2013materials}. ALD forms oxide layers by building up the oxide sequentially, whereas electron beam evaporation 
produces oxides in a single deposition step.

The inductive element is a single U-shaped inductor made of TiN, which couples to a feedline with a coupling quality factor $Q_c \approx$ 6500. 
Due to the TiN’s higher kinetic inductance (189 pH/$\square$ $\pm$ 3 pH/$\square$) compared to aluminum, the inductive element is shorter than an equivalent Al inductor. An equivalent Al inductor would need to be longer and more meandered, which would increase parasitic capacitance and reduce the participation of the trilayer capacitors. 
The Al layer also forms the ground plane, upper capacitive electrodes, and bonding pads. To mitigate flux noise, the ground plane was perforated. 
The trilayer design minimizes electromagnetic coupling to stray noise sources and concentrates the electric field within the small oxide volume 
(5.3 $\mu$m $\times$ 5.3 $\mu$m $\times$ 49 nm) [see Fig. \ref{fig:Device_image}(c,d)]. This concentrated electric field allows us to strongly couple to the TLSs 
and the majority of loss originates from TLSs within the oxide barrier \cite{muller2019}. Additional fabrication details are provided in 
Supplemental Material I.

\subsection{Device Operation}\label{Device Operation}

Initial measurements of the oscillator were performed in the first dilution refrigerator, which we designate as the ‘control’ condition. After approximately five 
months, the device was transferred to a second dilution refrigerator before further experiments were conducted. We refer to this period as the ‘control-aged’ stage to 
account for any potential aging effects on device parameters. Aside from relocating the device, all other experimental conditions and setup parameters 
in the second dilution refrigerator remained unchanged for the remainder of the study.

The bias line ($V_g$) is placed in a virtual ground between the capacitors, minimizing coupling to the rf mode [Fig. \ref{fig:Device_image}(a,b)]. The voltage bias generates a uniform electric field $E_g = V_g/(2d_0)$ across the bias bridge, 
where $d_0$ is the dielectric thickness. The capacitors were designed with a small volume to achieve 
strong coupling between TLSs and the oscillator \cite{sarabi2015}.

The rms electric field in the capacitors is calculated as 
$E_{\rm{rms}} = \sqrt{\hbar \omega_c / 2 \epsilon V_{T}}$, where $V_{T}$ 
is the total capacitor volume and $\omega_c$ is the oscillator frequency \cite{sarabi2015}. The 
dielectric constant of the oxide is $\epsilon = \epsilon_r \epsilon_0$, 
with $\epsilon_r \approx 10$. The capacitor volume, $V_{T}$, in this study is 5.6 $\mu \rm{m}^3$, significantly 
smaller than the 78 $\mu \rm{m}^3$ reported in \cite{sarabi2016}, resulting in an $E_{\rm{rms}}$ of 
53 V/m, approximately 2.5$\times$ stronger than previous work.

\section{Results}\label{Results}

\subsection{TLS spectroscopy}\label{TLS Spectroscopy}

\begin{figure}[ht!]
    \centering
    \includegraphics[width=3.3in]{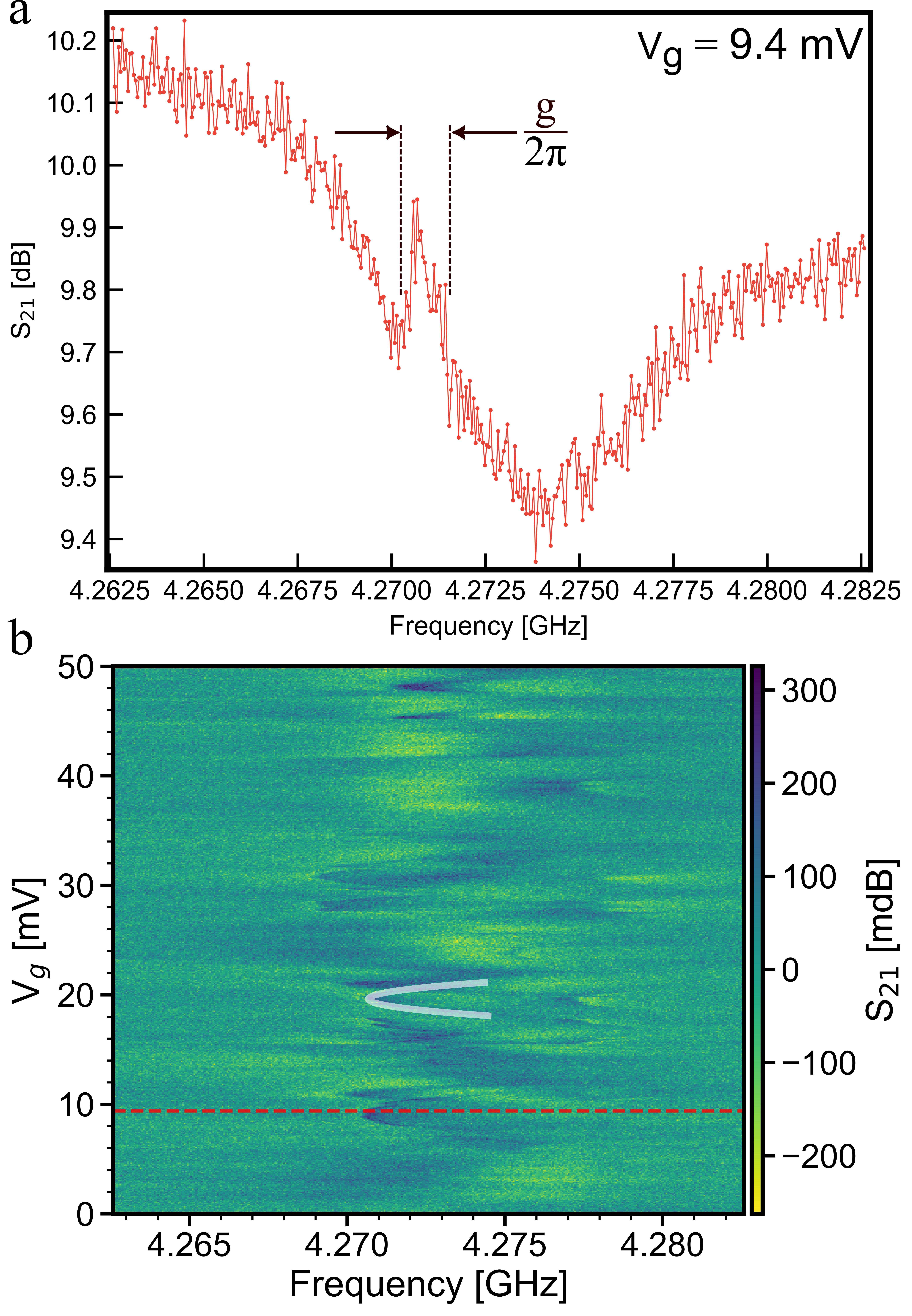}
    \caption{
        (a) Transmission (S$_{21}$) of a oscillator interacting with a TLS at $V_g$ = 9.4 mV. 
        The width of the TLS-induced avoided crossing is highlighted with black dashed lines, which is the outcome of the oscillator's 
        coupling to the TLS. The TLS-oscillator coupling strength, $g/2\pi$, is calculated to be 1.3 MHz which matches the width of the avoided-crossing.
        (b) Transmission (S$_{21}$) as a function of applied voltage ($V_g$) and oscillator frequency. The TLS-oscillator interactions appear as hyperbolic shapes following Eq. \ref{eq1:TLS_energy}. 
        An example is shown with the white curve corresponding to an extracted dipole moment $p_z$ = 0.24 e$\rm{\AA}$. The individual Lorentzian spectrum shown in (a) is a line cut at the red dashed line. Data processing for this figure is described in Supplemental Material III.}
    \label{fig:ABAA_before_after}
\end{figure}

To quantify TLS-oscillator interactions, we measured the transmission spectrum (S$_{21}$), 
as shown in Fig. \ref{fig:ABAA_before_after}. In some spectra, peaks in the Lorentzian profile 
indicate interference caused by TLS coupling to the oscillator [Fig. \ref{fig:ABAA_before_after}(a)] 
\cite{sarabi2016,kristen2024,sarabi2015,deGraff2021,hung2022}. To isolate TLS-induced avoided crossings, 
we averaged out the oscillator’s Lorentzian profile and time-varying components along the voltage axis 
(see Supplemental Material III). 

The applied electric field bias $E_g$ modifies the TLS energy according to Eq. \ref{eq1:TLS_energy} 
\begin{equation} 
    \epsilon = ({\Delta_0}^2 + ({\Delta - 2\vec{p} \cdot \vec{E_g}})^2)^{1/2}, \label{eq1:TLS_energy} 
\end{equation} 
where $\Delta_0$ represents the tunneling barrier height, $\Delta$ is the TLS asymmetry energy, 
and $\vec{p}$ is the TLS dipole moment 
\cite{anderson1972,phillips1972,sarabi2016,hung2022}.

Figure \ref{fig:ABAA_before_after}(b) is a measurement of the oscillator spectrum as the TLS energy is tuned via the applied voltage bias 
$V_g$. Each TLS spectrum required approximately 50 hours to complete, with measurements performed with an intermediate frequency bandwidth 
of 1 Hz to ensure sufficient averaging of avoided-level crossings. These measurements, conducted prior to any treatment steps 
and in the second dilution refrigerator (see Section \ref{Methods}A), serve as the control for the remainder of this study. All measurements, 
including those depicted in Fig. \ref{fig:ABAA_before_after}, were conducted at the single-photon 
level \cite{sarabi2014cavity}. 

\begin{figure}[ht!]
    \centering
    \includegraphics[width=2.5in]{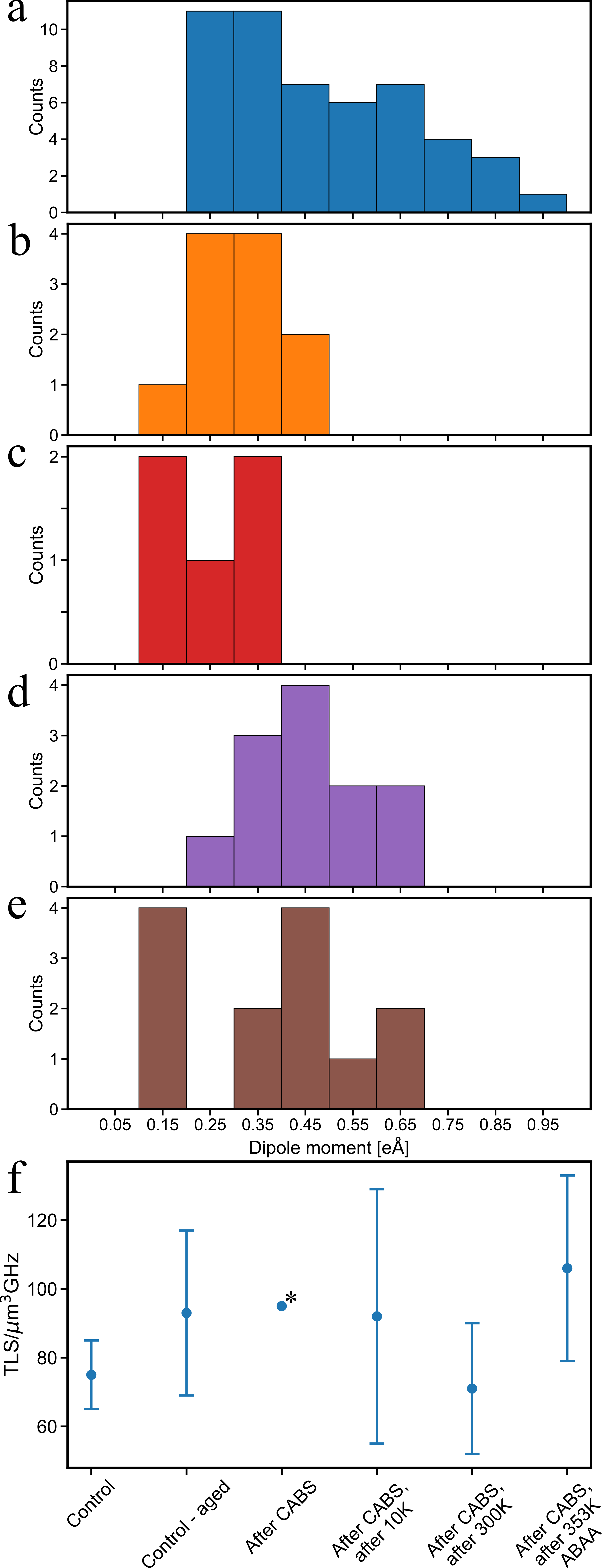}
    \caption{Extracted dipole moments for TLS spectra collected after different treatments. The histograms represent the following: (a) control (blue, initial measurement), (b) control-aged, (orange, after aging 5 months),  
    (c) 10 K temperature cycle after CABS (red), (d) 300 K temperature cycle after CABS (purple), and (e) after a 353 K alternating bias assisted annealing treatment (brown). We don't present any data on electric 
    dipole moments directly after CABS because there were no TLS-induced hyperbolas found in this scan.
    (f) The extracted TLS densities for the different processing steps. $^*$(Calculated value of TLS density from the main paper after CABS treatment.)}
    \label{fig:Pre_post_ABAA_histogram}
\end{figure}

To analyze the TLS spectra over a large voltage range, image processing and a machine learning algorithm were employed to 
identify individual hyperbolic curves. To optimize the fit to the TLS spectrum, we first implemented a denoising step and a 
gradient edge filtering scheme to enhance the definition of TLS avoided crossings. The machine learning algorithm then identified 
TLS hyperbola features, which were fit using Eq. \ref{eq1:TLS_energy}. Dipole densities were calculated from the extracted TLS dipole 
moment statistics and compared across different device treatments. Additional details on the TLS fitting procedure are provided in 
Supplemental Material III.
\begin{center}
\begin{table*}[ht!]
\caption{Table of average TLS electric dipole moments ($p_z$), TLS densities, low-power loss tangent values, 
        and critical photon number for different in-situ and thermal treatments. Two loss tangents are provided, one measured from power saturation and the other calculated from TLS density using Eq. A13 \cite{hung2022}. Errors represent one standard deviation from the mean. 
        $^{\dag}$(Calculated from the analysis of avoided-level crossings in Fig. \ref{fig:After_CABS_closer_look}). *(not measured)}
\begin{ruledtabular}
\begin{tabular}{c c c c c c}
Treatment & mean $p_z$ [e$\rm{\AA}$] & TLS density [$\frac{\rm{TLS}}{\mu m^3 \rm{GHz}}$] &  $\tan\delta^0_{\rm{TLS}}$ [$\times$ 10$^{-3}$] & $\tan\delta^{0calc}_{\mathrm{TLS}}$ [$\times$ 10$^{-3}$] & $n_c$  [$\times$ 10$^{-3}$] \\ [0.5ex] 
\hline
Control & 0.49 $\pm$ 0.03 & 75 $\pm$ 10 &  1.89 $\pm$ 0.03 & 0.21 $\pm$ 0.03 & N/A \\
\makecell{Control-aged \\ (Aged 5 months)} & 0.31 $\pm$ 0.02 & 93 $\pm$ 24 &  *  & 0.12 $\pm$ 0.03  & * \\
\makecell{CABS} & N/A & 95$^{\dag}$ &  2.00 $\pm$ 0.03 &  N/A  & 22 $\pm$ 2 \\
\makecell{10 K thermal cycle} & 0.24 $\pm$ 0.04 & 92 $\pm$ 37 &  2.00 $\pm$ 0.05 & 0.07 $\pm$ 0.03 & 23 $\pm$ 2 \\
\makecell{300 K thermal cycle} & 0.45 $\pm$ 0.04 & 71 $\pm$ 19 &  * & 0.14 $\pm$ 0.04 & * \\
\makecell{Alternating $V_g$ at 353 K} & 0.38 $\pm$ 0.05 & 106 $\pm$ 27 & 1.64 $\pm$ 0.03 & 0.17 $\pm$ 0.04 & 17 $\pm$ 2 \\
\end{tabular}
\end{ruledtabular}
\label{tabdipoledens}
\end{table*}
\end{center}

We extracted TLS hyperbolas from the control with an average dipole moment of 0.49 e$\rm{\AA}$ [Fig. \ref{fig:Pre_post_ABAA_histogram}(a)] and a TLS density of  75 $\pm$ 10 $\rm{TLS}/(\mu m^3\rm{GHz})$. 
After the device was aged for five months, the average TLS dipole moment dropped to 0.31 $\pm$ 0.02 e\AA, 
but the density is similar, 93 $\pm$ 24 $\rm{TLS}/(\mu m^3\rm{GHz})$ [see Fig. \ref{fig:Pre_post_ABAA_histogram}(b)]. Statistical analysis of the control experiment  
confirmed that our results are consistent with previous studies on TLS dipole moments and densities
in oxide films \cite{hung2022,sarabi2016}.

The TLS-oscillator coupling strength, $g$~=~$p_z \sqrt{\omega_c/2\epsilon \hbar V_{T}}$
can be calculated using the extracted dipole moment \cite{sarabi2015}. Here, $p_z$ is the projected dipole moment along the 
applied $E_{g}$ direction. Using the average dipole moment of the control $p_z = 0.5$ e$\rm{\AA}$, 
the TLS coupling strength is calculated as $g/2\pi = 1.3$ MHz, which matches the width in Fig. \ref{fig:ABAA_before_after}(a).

\subsection{Alternating Bias Treatments}\label{Application of Alternating Bias Treatments}

To investigate the effects of alternating bias on TLSs at cryogenic temperatures, we applied a time-varying bias to the voltage gate 
following procedures described in \cite{pappas2024alternating,wang2024precision}. Alternating pulses of $\pm$10 V 
were applied for 10 s each, repeated 2800 times over 30 hours. Each pulse was separated by a 10 s dwell
time at 0 V, creating an electric field of $E_g = \pm 100$ MV/m across the oxide [Fig. \ref{fig:Device_image}(d)]. 

\begin{figure*}[ht!]
    \centering
    \includegraphics[width=7in]{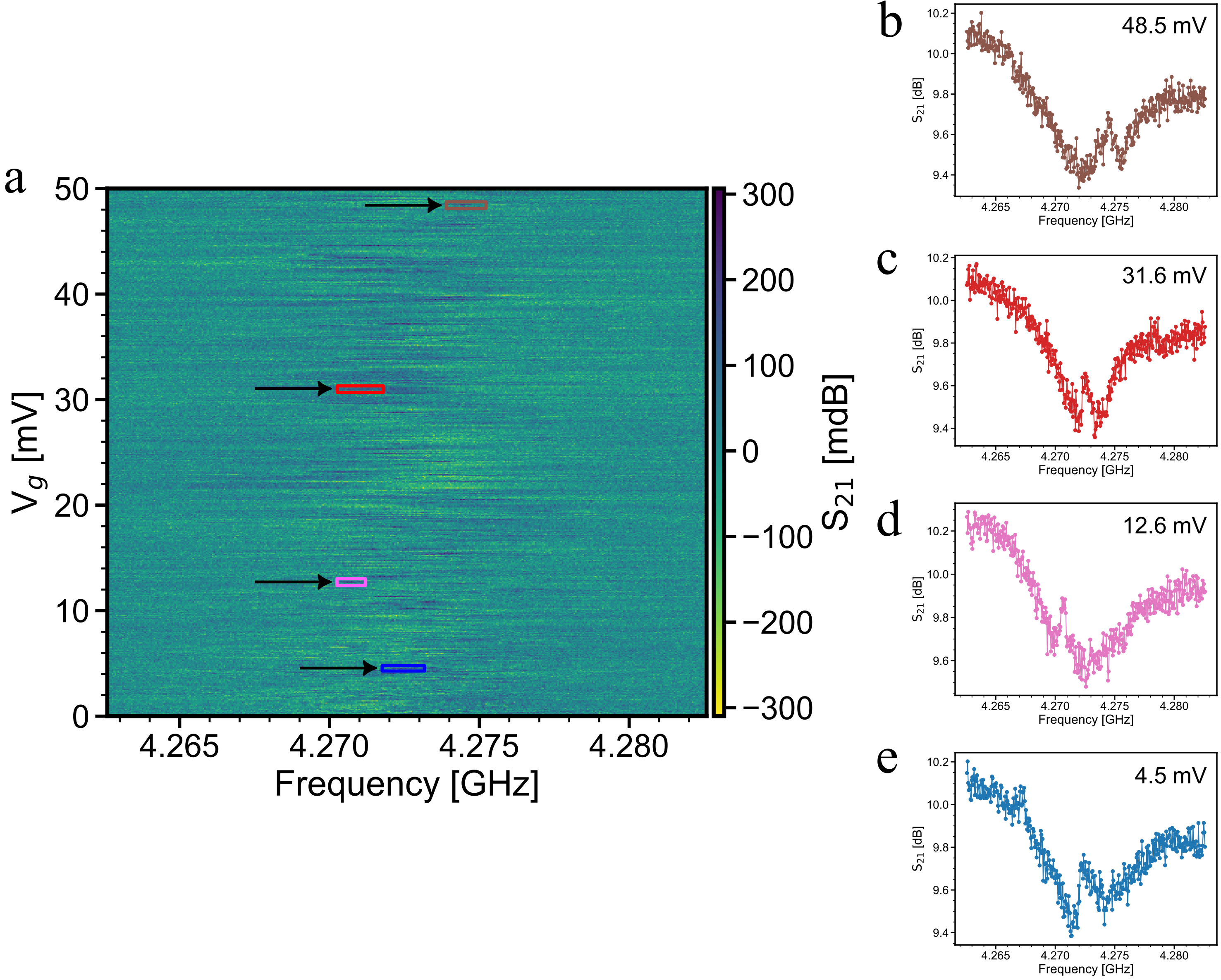} 
    \caption{
    (a) The transmission (S$_{21}$) versus applied voltage ($V_g$) and oscillator frequency spectrum after the CABS treatment 
    reveals sharp features during the voltage scan. Post-CABS treatment reveals no TLS-induced hyperbolas. Instead, the TLS 
    spectrum appears to exhibit frequency shifts, with transient avoided crossings lasting less than 5 minutes. Data processing for this figure is described in Supplemental Material III. (b-e) show individual tranmission spectra at the points indicated in (a). Avoided crossing with a width of approximately 1 MHz can be observed.
    }
    \label{fig:After_CABS_closer_look}
\end{figure*}

Following the treatment, TLS spectroscopy was conducted over a voltage range of 0–50 mV with 100 $\mu$V steps, taking 
a total of 50 hours. The results, shown in Fig. \ref{fig:After_CABS_closer_look}(a) do not present any TLS-induced hyperbolas.  
However, Fig. \ref{fig:After_CABS_closer_look}(a) reveals sharp, unidentified features 
in the TLS spectra. Figure \ref{fig:After_CABS_closer_look}(b-e) are linecuts of Fig. \ref{fig:After_CABS_closer_look}(a) at specific bias points.

Closer inspection of the resonator transmission reveals avoided-crossings with 
magnitudes similar to those observed in the control spectra (Fig. \ref{fig:ABAA_before_after}).
The estimated coupling strength ($g/2\pi$) associated with these 
features is approximately 1 MHz. However, these avoided crossings are highly transient, lasting less than 
5 minutes. Since each scan takes 6 minutes this measurement only provides an upper bound on the TLS frequency jitter. The next curve typically does not capture a TLS at the same frequency.

Despite Fig. \ref{fig:After_CABS_closer_look} not presenting hyperbolas we were able to extract approximate TLS parameters using the avoided-crossing width as we did in Fig. \ref{fig:ABAA_before_after}(a) (see Table \ref{tabdipoledens}).
In total, 122 individual avoided crossings were found within a 12 MHz bandwidth of the oscillator, out of 500 different 
applied bias points. Electric dipole moments could not be fit to these features because full hyperbolas were not observed. Instead, we compared the CABS sample 
and the control sample using the ratio of avoided crossings to total bias points. One interpretation 
of the CABS spectrum is that, at each instant of an avoided crossing, a different resolvable TLS hyperbola, statistically independent from others, would be observed if the 
measurement were instantaneous. With this assumption, we measured 122 TLS in 500 curves, or a probability of 0.24 TLS per transmission spectrum. Before CABS the measurements are not statistically independent. A typical TLS hyperbola has a width of 0.75 mV and we measured approximately 50 TLS over 200 mV. Therefore, at a random bias there are 0.19 TLS per transmission spectrum.
 
Although the TLS landscape is significantly altered by the CABS procedure, the density of avoided crossings throughout the bias sweep seems to be comparable.  Assuming the same dipole moment distribution as the control sample, the TLS density of the CABS sample would be $95~\rm{TLS}/(\mu m^3\rm{GHz})$, 
similar to the control sample (Eq. A14, Ref. \cite{hung2022}).

\begin{figure}[ht!] 
    \centering
    \includegraphics[width=3.4in]{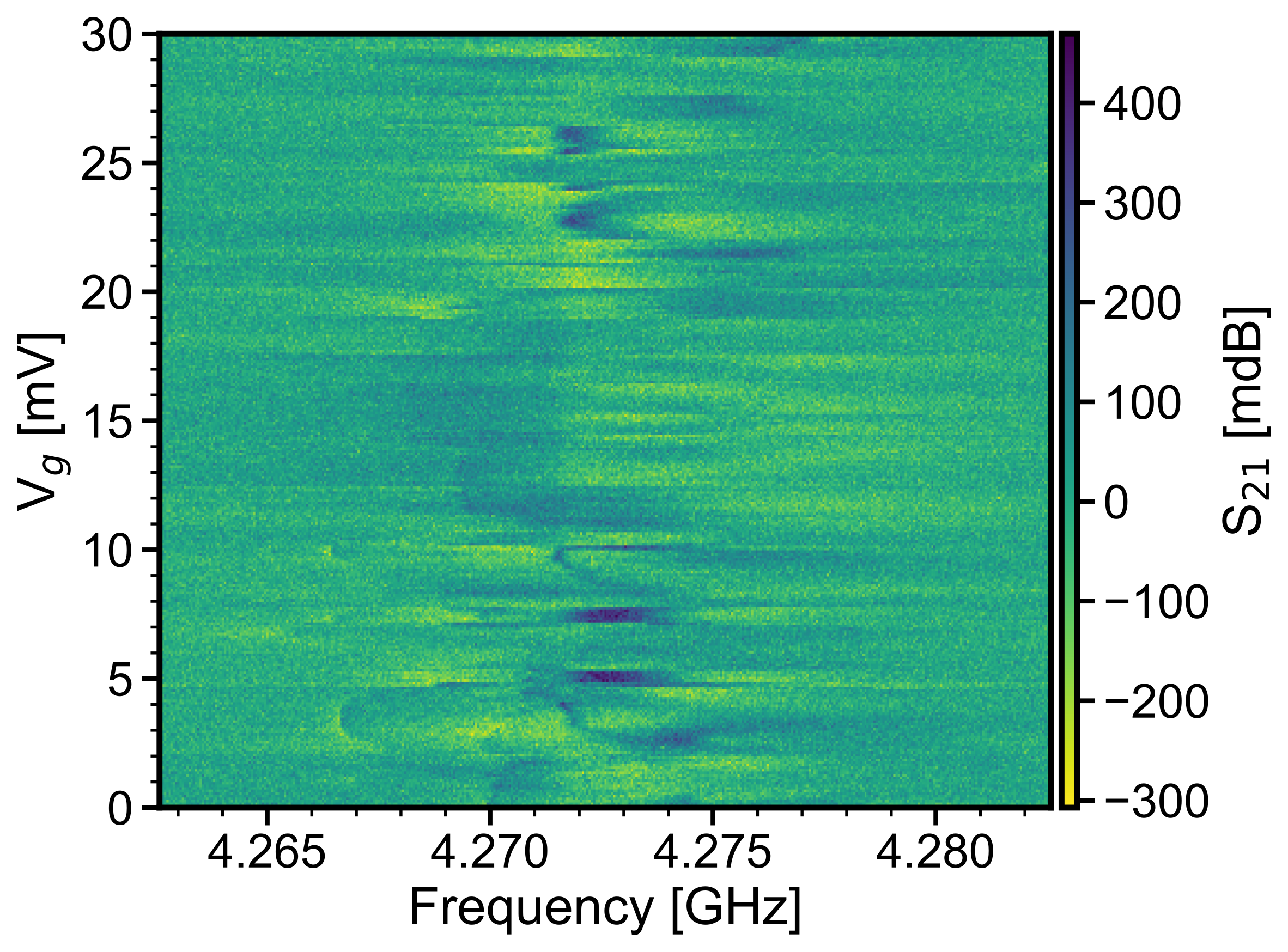} 
    \caption{
    Measured transmission spectrum (S$_{21}$) after thermal cycling to 10 K. Hyperbolas can be seen in this spectrum. Data processing for this figure is described in Supplemental Material III.
    }
    \label{fig:After_CABS_and_2nd_tempSweep}
\end{figure}

\subsection{Thermal cycling}\label{Thermal cycling}

Following the CABS treatment, we investigated the effects of thermal cycling to 10 K.
The spectroscopy is shown in Fig. \ref{fig:After_CABS_and_2nd_tempSweep}. 
After temperature cycling we measured a density of 92~$\pm$~37~$\rm{TLS}/(\mu m^3\rm{GHz})$, 
[Fig. \ref{fig:Pre_post_ABAA_histogram}(c)].

Similarly, after cycling to 300 K (room temperature), the TLS density and average electric dipole moment had values 
similar to those observed in the pre-CABS state [see Table \ref{tabdipoledens} 
and Fig. \ref{fig:Pre_post_ABAA_histogram}(d)]. 

We next explored the influence of TLS density using a non-cryogenic alternating voltage bias treatment \cite{pappas2024alternating,wang2024precision}, 
performed on a probe station for 30 hours. During this test, the platform supporting the device was heated to 353 K for the entire duration. 
A $\pm$10 V alternating bias was applied with 1-second pulses to minimize the risk of dielectric breakdown. TLS spectroscopy conducted 
after this treatment revealed a density of 106~$\pm$~27~$\rm{TLS}/(\mu m^3\rm{GHz})$.

Figure \ref{fig:Pre_post_ABAA_histogram}(e) presents a summary of dipole moment distributions, and Table \ref{tabdipoledens} lists densities. The average electric dipole moment after the 353 K alternating bias treatment was 0.38 $\pm$ 0.05 e\AA. 
This electric dipole moment was similar to the average dipole moment in the initial control and after the 300 K cycle 
measurement. No matter the treatment we applied all the densities were very similar. 

\subsection{TLS low power saturation}\label{TLS low power saturation} 

To further understand the effects of CABS on TLSs, we studied the power dependence of TLS loss 
\cite{gao2008physics}. Figure \ref{fig:power_sweep} shows the loss tangent, $\tan\delta$, as a function 
of the microwave energy injected into the resonator. Measurements were performed at each stage of sample 
processing: control, CABS, 10 K thermal cycle, and 353 K alternating bias 
treatment \cite{pappas2024alternating,wang2024precision}. The control measurement was conducted 
on the same sample, but in a different cryostat.

The loss tangent curve was fitted using Eq. \ref{eq1:TLS_pwr_dependence} 
\begin{equation} 
    \tan\delta = \frac{\tan\delta^0_{\mathrm{TLS}}}{\sqrt{1+\frac{n}{n_c}}} + \tan\delta_E \label{eq1:TLS_pwr_dependence},
\end{equation} 
where $n$ and $n_c$ are the average photon number and critical photon number, respectively, and $\tan\delta_E$ is the 
power-independent contribution to the loss tangent. From the standard tunneling model, the low-power loss tangent due to 
TLSs is $\tan\delta^0_{\mathrm{TLS}}$ = $\pi P_0|\vec{p}|^2/3 \epsilon$, 
where $\epsilon$ is the dielectric permittivity of the oxide layer and $P_{0}$ is the TLS 
density \cite{phillips1987,muller2019}. The average $\tan\delta^0_{\mathrm{TLS}}$ was calculated 
as $(1.88 \pm 0.07)\times10^{-3}$ (black dashed line in Fig. \ref{fig:power_sweep}), and was consistent across all processing steps 
(see Table \ref{tabdipoledens}). This value aligns 
with previous measurements for alumina, where $\alpha$-Al$_2$O$_3$ exhibited loss tangents of 1.50$\times$10$^{-3}$ and $\alpha$-AlO$_3$ 
exhibited 0.98$\times$10$^{-3}$ \cite{hung2022}.

\section{Discussion}\label{Discussion}

\begin{figure}[ht!]
    \centering
    \includegraphics[width=3.4in]{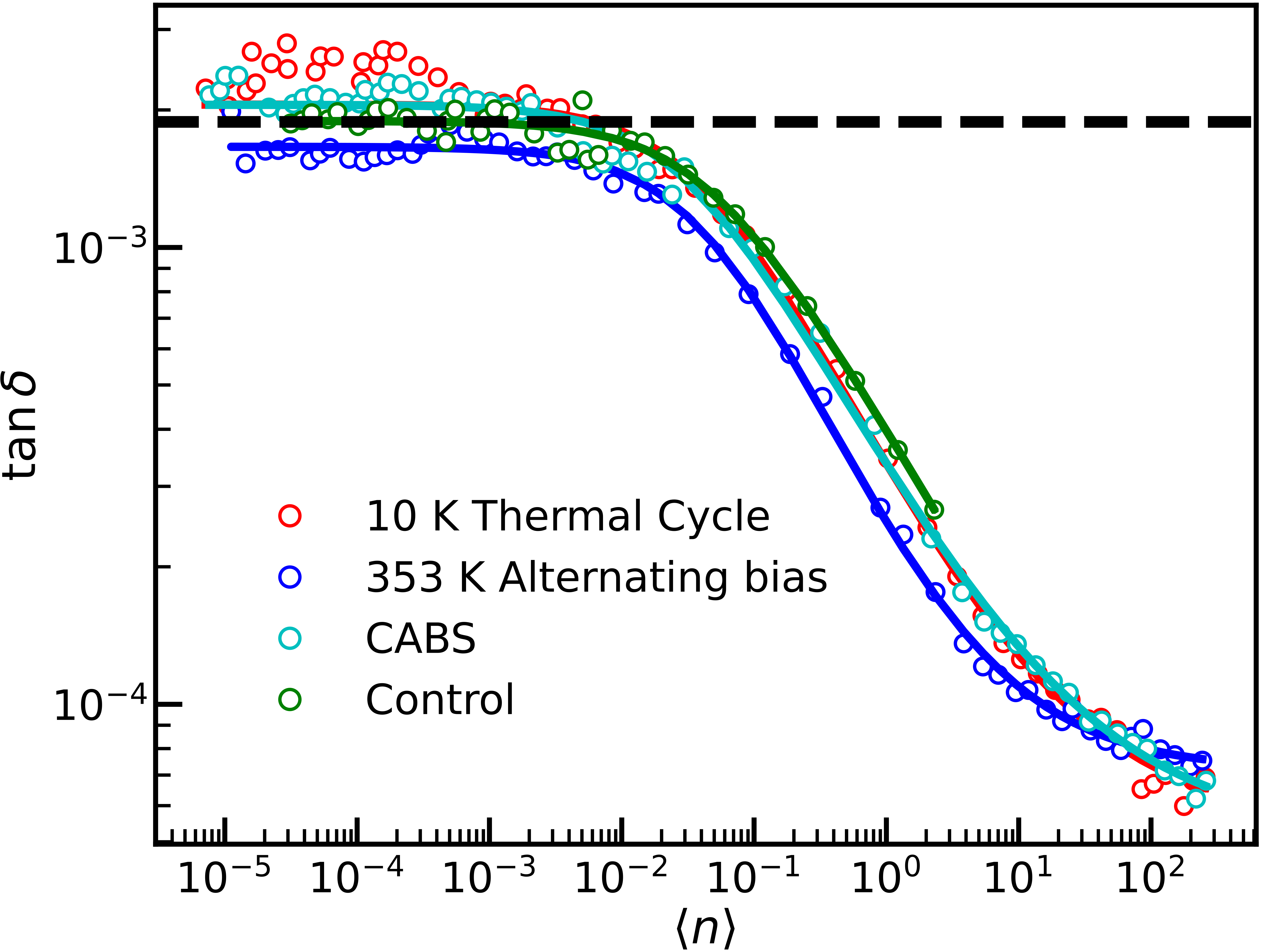}
    \caption{
        Loss tangent ($\tan\delta$) versus average photon number ($\braket{n}$) under various experimental conditions. 
        Loss tangent curves are shown for the control (green circles), after CABS 
        treatment (teal circles), after 10 K thermal cycling (red circles), and after alternating voltage biasing annealing at 353 K
        (blue circles). All measurements were conducted at a base temperature of 10 mK.
        Solid lines represent fits to the loss tangent data using Eq. \ref{eq1:TLS_pwr_dependence}, with colors corresponding to each 
        treatment step. The black dashed line indicates the average intrinsic TLS loss ($\tan\delta^0_{\mathrm{TLS}}$ = 1.88$\times$10$^{-3}$), which provides a baseline for comparison.}
    \label{fig:power_sweep}
\end{figure}

Based on the density of TLS-induced hyperbolas in Fig. \ref{fig:ABAA_before_after}(b),
the contribution of strongly coupled TLSs to the loss tangent is calculated to be 
$\tan\delta^0_{\mathrm{TLS}}$ = 0.12$\times10^{-3}$, which is an order of magnitude 
lower than the fitted value in Fig. \ref{fig:power_sweep}. The calculation of $\tan\delta^0_{\mathrm{TLS}}$ was done using Eq. A13 in Ref. \cite{hung2022}. This calculated result for $\tan\delta^0_{\mathrm{TLS}}$ is on the order of the error margin 
of the fit (Fig. \ref{fig:power_sweep}), indicating that visible strongly coupled TLSs are a minor correction to the overall background 
loss. Additionally, following CABS and other treatments, the 
$\tan\delta^0_{\mathrm{TLS}}$ fits (Fig. \ref{fig:power_sweep}) converged to the same value, suggesting 
consistent behavior across the TLS population.

Our calculated TLS ac coupling, $g$/$2\pi$ = 1.3 MHz, agrees with the measured TLS-induced avoided-crossing width seen in Fig. \ref{fig:ABAA_before_after}(a) for the rms driven electric field of $E_{\rm{rms}}$ = 58 V/m in our capacitors. The driven field is calculated at the single photon regime.  Assuming that the 
$n_c$ extracted from Fig. \ref{fig:power_sweep} corresponds to strongly coupled TLSs, the TLS relaxation 
rate $T^{\rm{TLS}}_1$ is approximately 1 $\mu$s when we use our measured $n_c$ = 21$\times$10$^{-3}$, $p_z$ = 0.5 e$\rm{\AA}$, and $2\pi$$g$. This $T^{\rm{TLS}}_1$ agrees with previous 
studies of strongly coupled TLSs 
\cite{lisenfeld2016decoherence,khalil2014landau,lisenfeld2010measuring,cho2023simulating}. From our measured TLS ac coupling and device statistics we show consistency with other TLS measurements of amorphous oxides. 

The initial TLS spectroscopy revealed TLS-induced hyperbolas (Fig. \ref{fig:ABAA_before_after}), consistent with previous reports \cite{hung2022,sarabi2016,sarabi2015,de2021quantifying}. These features persisted for several hours, in agreement with observations in superconducting qubits \cite{klimov2018}. Following this initial measurement, we applied CABS to the dielectric film (see Results for details). The application of CABS follows a protocol similar to that in Refs. \cite{pappas2024alternating,wang2024precision}. After CABS, the TLS-induced hyperbolas were absent and could not be fit to Eq. \ref{eq1:TLS_energy}. Instead, the spectroscopy displayed randomly distributed, sharp, peaks [see Fig. \ref{fig:After_CABS_closer_look} (b-e)]. These avoided crossings were highly transient, persisting for less than 5 minutes, and subsequent scans did not detect a TLS at the same frequency. Since each scan took 6 minutes, this measurement provides an upper bound on the TLS frequency jitter.

After CABS, the calculated ratio of avoided crossings to total bias points was within a factor of 1.3, as determined by 
individually counting TLSs with coupling strengths ($g/2\pi$) and avoided crossing depths comparable to those shown in
Fig. \ref{fig:ABAA_before_after}(a) (see Results). If each avoided-level crossing corresponds to a distinct TLS, the estimated 
density is approximately 95 $\rm{TLS}/(\mu m^3\rm{GHz})$. These results indicate that, although hyperbolic features are absent 
from the TLS spectrum after CABS, the overall TLS distribution likely remains unchanged.

Strongly coupled TLSs reappeared after thermal cycling to 10 K and subsequent cooling to 10 mK 
(Fig. \ref{fig:After_CABS_and_2nd_tempSweep}), with extracted dipole moments ranging from 0.5 
to 0.35 e$\rm{\AA}$. The 10 K thermal cycling experiment exhibited a TLS density comparable 
to the values observed in the control experiment. These results are consistent with previous 
experimental observations, such as the breakdown of the TLS model above 15 K in sound speed 
measurements for $\alpha$-GeO$_2$ and $\alpha$-B$_2$O$_3$ 
\cite{rau1995acoustic}, which indicated temperature-dependent changes in TLS behavior.
There has also been a recent study by Zanuz et al. that showed that TLS defect modes 
in qubits rearranged after thermal cycling to room temperature \cite{colao2025mitigating}. Our 
findings complement these prior works by demonstrating the reversible nature of strongly coupled 
TLSs in oxide films at temperatures above 10 K. The reappearance of TLSs after thermal cycling 
suggests that the TLS ensemble undergoes structural or energetic reorganization at elevated 
temperatures. These results indicate a dielectric transition associated with overcoming the TLS barrier height.

Random fluctuations in the frequency of the TLS-induced avoided crossing after CABS (Fig. \ref{fig:After_CABS_closer_look}) 
may suggest the presence of temperature-dependent, incoherent two-level systems known as two-level fluctuators (TLFs). TLFs 
are characterized by rapid, telegraph-like switching between two states, which can significantly shift the TLS 
frequency \cite{meissner2018probing, de2021quantifying}. While TLFs couple dispersively to resonant TLSs, they do not 
contribute to the total intrinsic loss. Instead, TLFs shift the TLS frequency in response to local temperature fluctuations. 
This mechanism is consistent with the observed behavior before CABS, where TLS frequency fluctuations occurred over 
timescales of hours \cite{klimov2018}. However, after CABS, the frequency fluctuations accelerated to the order of 
minutes, approximately 100 times faster, which is not consistent with typical TLF dynamics. This spectrum may therefore be related to 
multiple ensembles of TLFs changing after CABS.

Energy buildup during the implementation of CABS may influence the TLF ensemble or agitation of the TLS ensemble over time. 
One potential mechanism involves phonon bursts triggered by stress relaxation in the dielectric substrate. In superconducting devices 
used for dark matter detection and qubits, sudden bursts of phonons have been observed due to microfractures in the substrate caused by epoxy used 
to secure the devices \cite{aastrom2006fracture,yelton2025correlated,anthony2024stress,armengaud2016constraints}. During CABS, energy buildup 
can possibly occur during the alternating voltage sequence of 
$\pm$ 10 V ($\pm$ 100 MV/m). Alternating bias treatments have been shown to alter the atomic profile of the dielectric material 
\cite{pappas2024alternating}, potentially inducing piezo-tunneling effects in materials such as $\alpha$-Al$_2$O$_3$. 
The field strength used for CABS is of the same order of magnitude as the parasitic electric fields generated by trapped charges
in ALD-grown $\alpha$-Al$_2$O$_3$ junctions \cite{rafael2017piezo}.
These dielectric stresses may contribute to fluctuations in TLS populations.
These stress-induced relaxation events, known as phonon-only events, release energy randomly after CABS is applied, 
and gradually settle into equilibrium \cite{aastrom2006fracture,armengaud2016constraints,anthony2024stress,yelton2025correlated}. Such 
non-equilibrium energy release through phonon bursts could scramble TLF populations over time, disrupting the coherence of the TLS system.

\section{Conclusion}\label{Conclusion}

This study introduced Cryogenic Alternating Bias Stimulation (CABS) as an 
in-situ treatment that modifies the defect environment in a dielectric film.
Building on recent findings that alternating voltage biasing may indicate a structural change within qubits at 353 K \cite{pappas2024alternating}, 
we applied CABS for 30 hours and observed that TLSs are no longer stable in frequency. What remains in the spectrum after CABS are random avoided-level 
crossings that fluctuate on the order of minutes. 
One explanation is a buildup of non-equilibrium energy in the oxide film 
forming the capacitive elements of the oscillator that is gradually released scrambling the TLS stability. 

Our analysis following CABS processing suggest that the TLS density remained constant. Interestingly, thermal cycling above 10 K, we observed TLS frequency stability, demonstrating the reversible nature of Al$_2$O$_3$ films. The fact that CABS changes the TLS frequency can provide insight into the origins of why TLSs fluctuate.

\section{Acknowledgements}
We would like to acknowledge fruitful discussions with Moshe Schechter, Sergey Pereverzev, and Alexander Burin. In addition, we would like to give special 
thanks to David Pappas, who provided insight into the alternating bias technique. The authors gratefully acknowledge use of facilities and instrumentation in the UW-Madison Wisconsin Center for Nanoscale Technology. The Center  (wcnt.wisc.edu) is partially supported by the Wisconsin Materials Research Science and Engineering Center (NSF DMR-2309000) and the University of Wisconsin-Madison.
This research was supported by the U.S. Department of Energy, Basic Energy Sciences, under award DE-SC0020313. 
This work was performed under the auspices of the U.S. Department of Energy by Lawrence Livermore National Laboratory under contract DE-AC52-07NA27344. LLNL-JRNL-2011522.

\bibliographystyle{apsrev}
\bibliography{Blpper_citations} 

\widetext
\clearpage
\newpage
\begingroup
\begin{center}
\textbf{\large Supplemental Material for ``Non-equilibrium Dynamics of Two-level Systems directly after Cryogenic Alternating Bias"}
\end{center}
\endgroup

\def\thesection{\Roman{section}}

\setcounter{section}{0}
\setcounter{secnumdepth}{3}
\setcounter{equation}{0}
\setcounter{table}{0}
\setcounter{page}{1}
\setcounter{enumiv}{0} 

\setcounter{figure}{0} 
\renewcommand{\thefigure}{S\arabic{figure}} 
\renewcommand{\figurename}{Supplemental Fig}

\section{Fabrication details}\label{SupSec:Fabrication_details}
The devices are fabricated on high-resistivity intrinsic Si (100) wafers. Prior to fabrication, the wafers were cleaned with a preliminary piranha dip, followed by an RCA process, and then a final dilute HF dip (49 $\%$ HF diluted 50:1 with deionized water) to remove the native oxide. The oscillator’s base layer, serving as the inductor, consists of a 22 nm TiN film deposited by plasma-enhanced atomic layer deposition (ALD) with the substrate chuck maintained at 275$^{\circ}$C, using tetrakis(dimethylamido)titanium (TDMAT) at 75$^{\circ}$C and plasma N$_2$ as precursors. The kinetic inductance is calculated to be 189 pH/$\square$ $\pm$ 3 pH/$\square$, where the uncertainty is estimated from the frequency spread observed across devices. The base layer is patterned using photolithography with a Nikon i-line stepper and etched with a BCl$_3$/Cl$_2$ plasma.

A 49 nm Al$_2$O$_3$ dielectric layer was deposited by thermal ALD at a substrate temperature of 250$^{\circ}$C, using room-temperature precursors trimethylaluminum (TMA) and H$_2$O. This layer is patterned by photolithography and etched with a wet phosphoric acid etch (Transetch-N). The thickness uncertainty of the ALD-deposited layers is $\pm$ 4 nm, with an assumed film uniformity variation of $\sim$1$\%$.

The device’s top metallization, comprising the ground plane, bias lines, bond pads, and the capacitor’s upper electrode, is defined using image-reversal photoresist and the i-line stepper with a flood exposure. A 150 nm aluminum layer is deposited by e-beam evaporation and lifted off to complete the structure.

\section{Fridge and Measurement layout}\label{SupSec:Frige and measurement layout}

\begin{figure}[h!]
    \centering
    \includegraphics[width=6.1in]{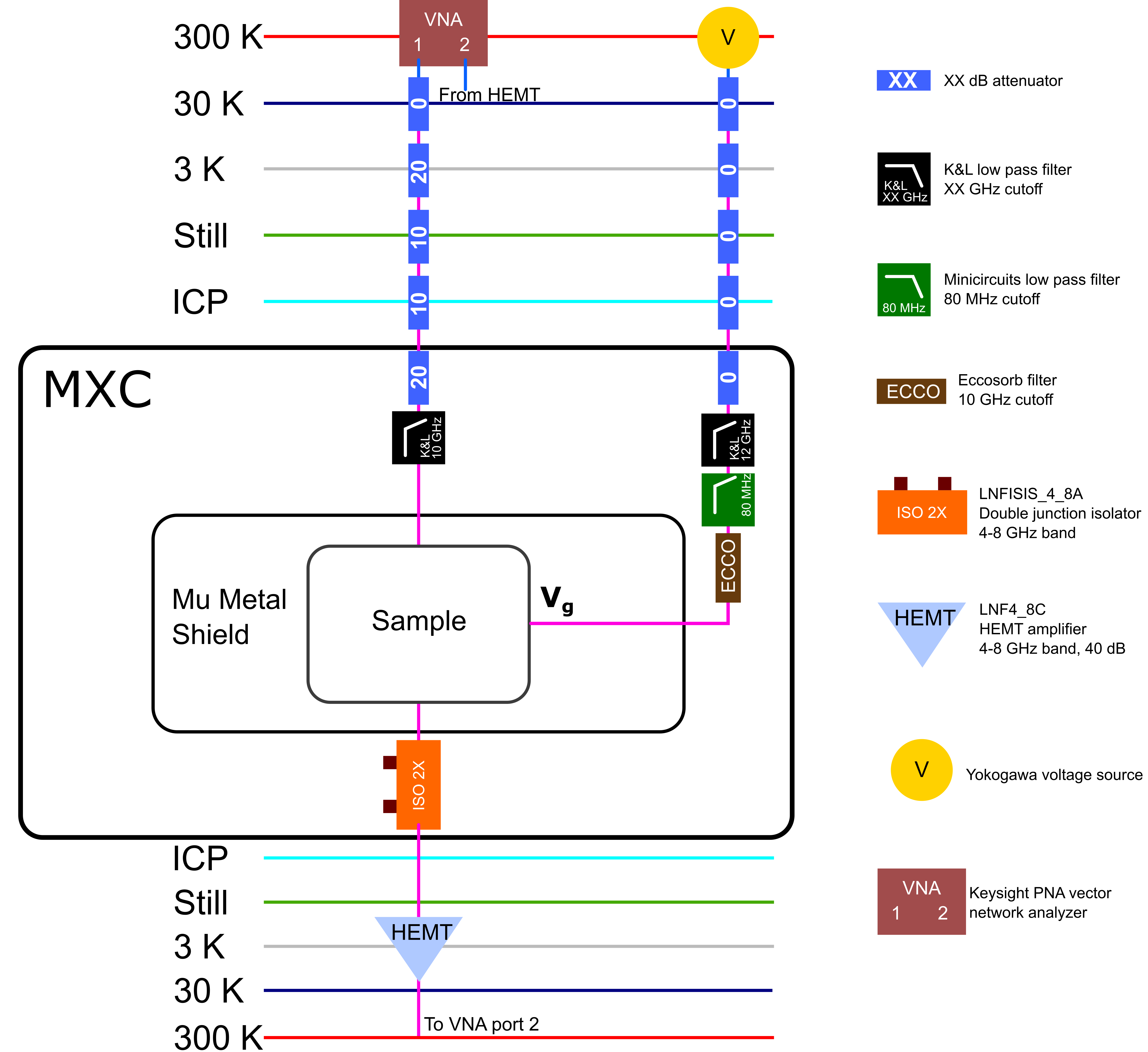}
    \caption{
        The wiring diagram illustrates the experimental setup for the oscillator device, highlighting key components 
        and their roles in ensuring optimal performance. The microwave signal path includes filters, isolators, and other 
        rf components, which are labeled on the right-hand side of the diagram. Attenuation values (indicated by blue squares) are specified 
        for each temperature stage, with a total attenuation of 60 dB. The sample is on a paddle mounted on the mixing chamber (MXC) stage, which is cooled to a base temperature of 10 mK using a dilution 
        refrigerator. Surrounding the MXC stage is a magnetic shield made of mu-metal, designed to attenuate stray magnetic 
        fields that could interfere with the device’s operation. Voltage dc biasing ($V_g$) is applied through a coaxial line with no attenuation.}
    \label{Supfig:Fridge_measurement_setup}
\end{figure}

To measure transmission ($\rm{S}_{21}$), we routed a microwave signal through a dilution refrigerator equipped with 
multiple stages of attenuation totaling 60 dB and a 10 GHz low-pass filter (see Supplemental Fig. \ref{Supfig:Fridge_measurement_setup}). 
The attenuation stages were used to minimize thermal noise. The device sample, measuring 8 mm $\times$ 8 mm, 
was mounted inside a gold-plated copper (Cu) enclosure to provide thermal and electrical stability. DC bias and microwave signals were 
delivered to the chip via SMP connectors and wire bonds, with GE varnish used to secure the device to the Cu housing.

Further down the signal chain, the microwave signal passed through two blocks of cryogenic isolators operating in the 4–8 GHz range 
to suppress back reflections and noise. A cryogenic amplifier (high electron mobility transistor, HEMT) amplified the signal by 
approximately 35 dB, ensuring sufficient signal strength for accurate measurement. The device was mounted on a paddle attached to the mixing chamber stage (MXC) of 
a Bluefors XLD-1000 dilution refrigerator, as shown in Supplemental Fig. \ref{Supfig:Fridge_measurement_setup}. The dilution refrigerator was operated at a base temperature of 10 mK.
Surrounding the paddle at the MXC was a magnetic shield made from mu-metal, which effectively blocked stray magnetic fields that could interfere with the device's operation.

To tune the energy of two-level systems (TLSs), we employed a voltage bias line with no attenuation. The bias line 
included two stages of low-pass filtering: an eccosorb filter to suppress high-frequency noise and a Minicircuits 80 MHz low-pass filter for additional noise 
reduction. The DC voltage bias was supplied using a Yokogawa GS200, providing precise and stable control over the applied voltage.

\section{TLS fitting details and ML algorithms}\label{SupSec:TLS_fitting_details}

\begin{figure}[ht!]
    \centering
    \includegraphics[width=5in]{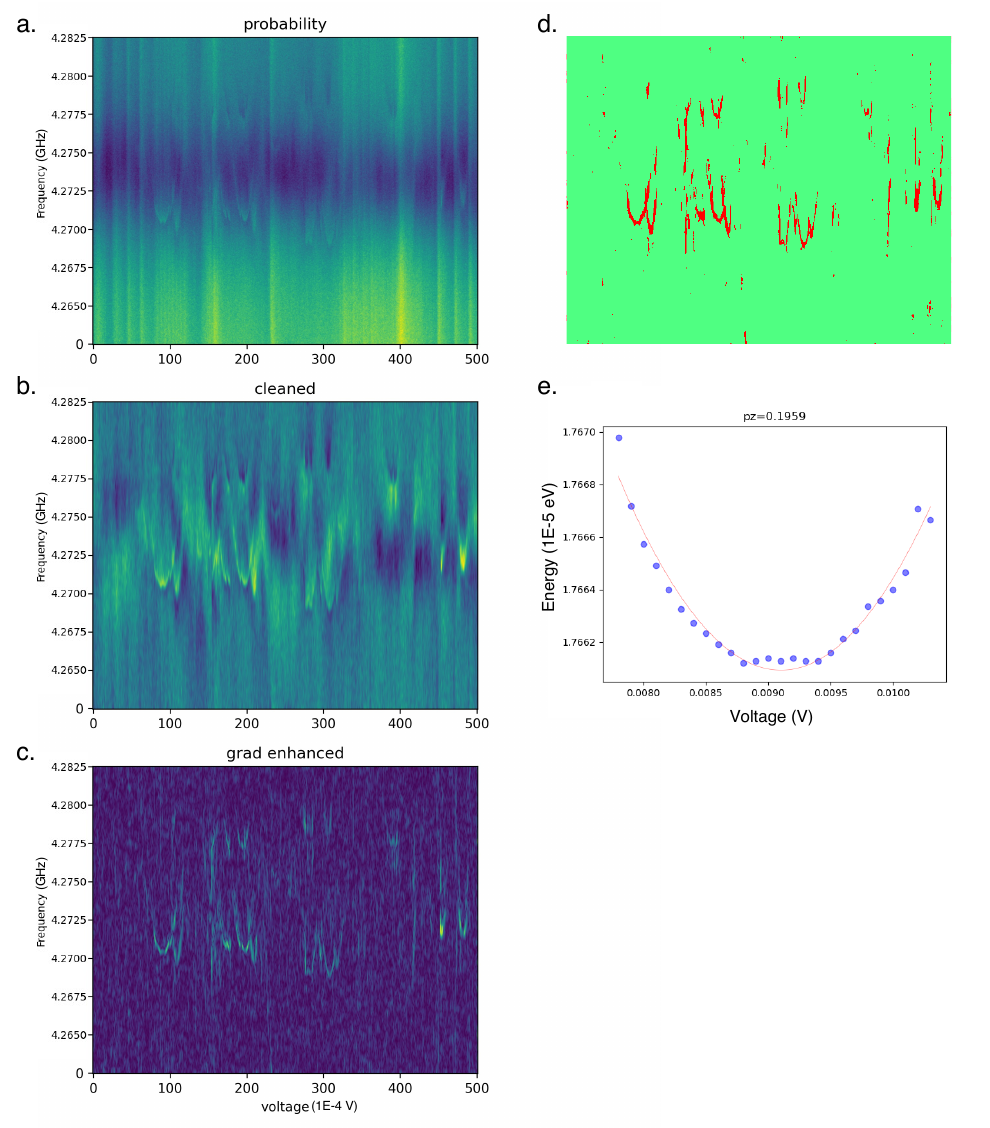}
    \caption{(a) Unprocessed experimental data. (b) Cleaned data, with fluctuations in time and in frequency removed as described in Supplemental Materials \ref{SupSec:TLS_fitting_details}. (c) Data after gradient enhancement. (d) Output of the ML model. 
    (e) Example of a fitted hyperbola, where the solid red line denotes the fit and the blue dots are taken from one hyperbola in (d).}
    \label{Supfig:gradient_and_ML_illustration}
\end{figure}

Transmission spectra (S$_{21}$) plots were processed using several steps before fitting the TLS energy vs. applied electric bias hyperbolas, Eq. \ref{eq1:TLS_energy} from the main paper, to features in those plots. These steps are summarized below and the results are shown in Supplemental Fig. \ref{Supfig:gradient_and_ML_illustration}.  

The Lorentzian around the resonator frequency in the S$_{21}$ spectra was removed by subtracting the S$_{21}$ value at each frequency averaged over bias: $\tilde{P}(f,V) = P(f,V) - \bar{P}_{\text{bias}}(f)$. Similarly, the baseline shift over time was removed by subtracting the frequency-wise average:  $\tilde{P'}(f,V) = \tilde{P}(f,V) - \bar{P}_{\text{frequency}}(V)$. 
Then we applied a 1D gaussian filter from the scipy python package and the result of this step is shown in Supplemental Fig. \ref{Supfig:gradient_and_ML_illustration}(b).
The transmission spectra in the main paper and Supplemental Fig. \ref{Supfig:After_RT_cycle_and_ABAA} have only undergone the two averaging steps to visually depict changes in the TLS-induced hyperbolas.

We next calculate the magnitude of the gradient of $\tilde{P}'(n)$ with respect to the image array coordinates $(i, j)$, where $i$ and $j$ refer to bias and frequency:
$$
|\nabla \tilde{P}'(i,j)| = \sqrt{\left( \frac{\partial \tilde{P}'(i,j)}{\partial i} \right)^2 + \left( \frac{\partial \tilde{P}'(i,j)}{\partial j} \right)^2}
$$
By applying this technique, we enhanced the contrast of the TLS hyperbola traces in the cleaned image (Supplemental Fig. \ref{Supfig:gradient_and_ML_illustration}(c)).

The resulting image was processed using the Trainable Weka Segmentation plugin for the Fiji image processing package based on the ImageJ2 software. 
The machine learning (ML) model was trained by manually selecting the portions of the image that correspond to TLS hyperbolas and those that do not, by hand. 
The algorithm used in the Trainable Weka Segmentation plugin is the fast random forest classifier. The resulting classified image is shown in 
Supplemental Fig. \ref{Supfig:gradient_and_ML_illustration}(d). With this image we select a rectangular region for each hyperbola by hand and pixels in those 
regions are fit to Eq. \ref{eq1:TLS_energy} from the main paper in order to extract TLS parameters: $\Delta_0$, $\Delta$, and $p_z$ 
[Supplemental Fig. \ref{Supfig:gradient_and_ML_illustration}(e)]. The TLS densities were extracted from dipole moment histograms using the method 
detailed in Ref. \cite{hung2022}.

TLS hyperbolas are only counted if $\Delta_0$ is within the bandwidth, therefore the primary source of uncertainty in the 
TLS density is due to the number of TLS with $\Delta_0$ within the bandwidth sampled. We divided the bandwidth into bins and 
calculate the standard deviation of the populations of TLS with $\Delta_0$ in those bins. From this, we estimated the 
uncertainty in the total number of TLS with $\Delta_0$ in bandwidth by dividing that bin-wise standard deviation by the 
square root of the number of bins in bandwidth.

\section{TLS spectroscopic data and additional analysis}\label{SupSec:TLS_spec_additional_analysis}

\subsection{Post-room Temperature Thermal Cycle and Alternating Voltage Bias}\label{Post-room_temperature_thermal_cycle_and_ABAA}

\begin{figure*}[ht!]
    \centering
    \includegraphics[width=5in]{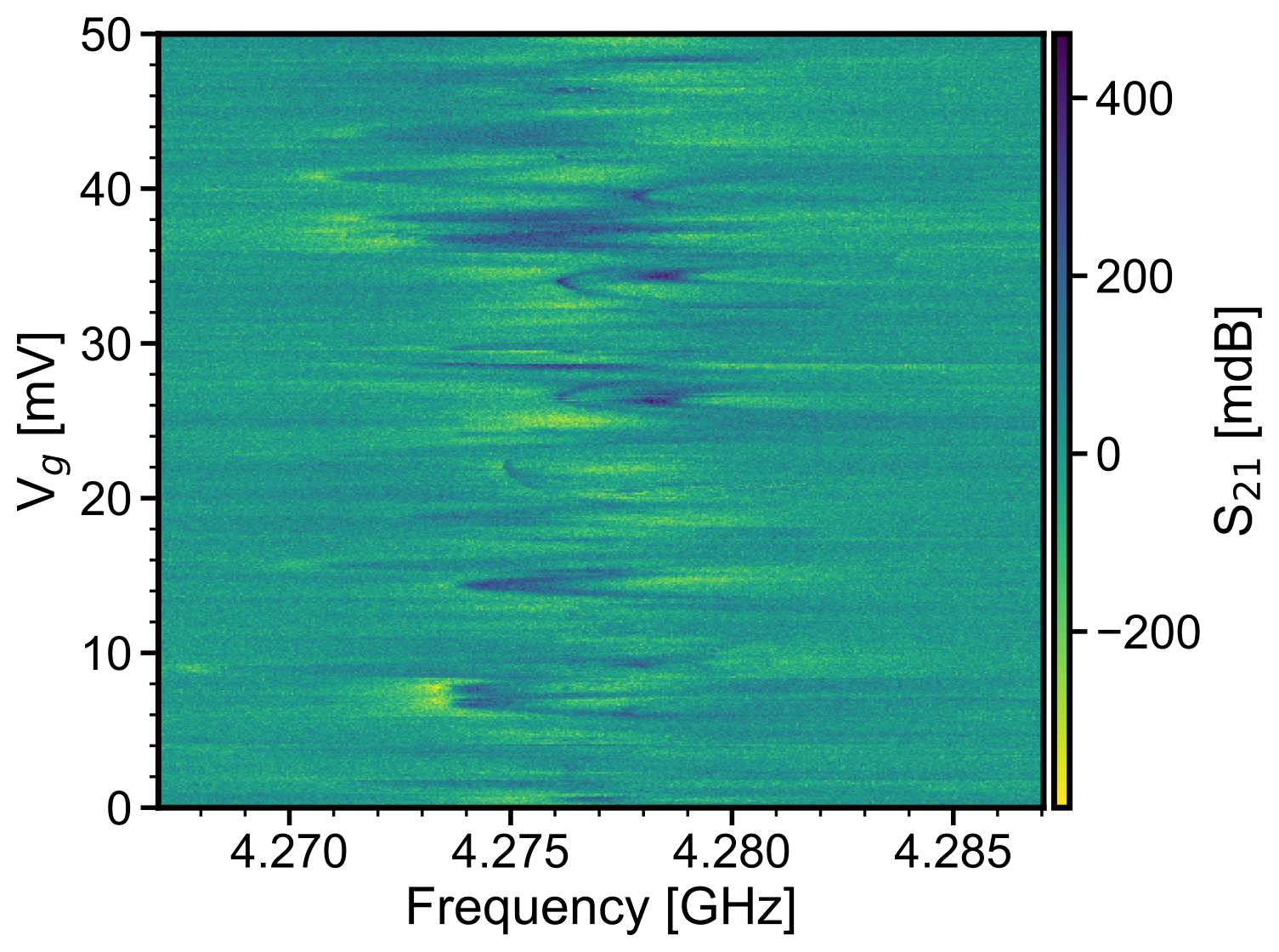}
    \caption{
    Transmission (S$_{21}$) versus applied voltage ($V_g$) and oscillator frequency was measured after alternating voltage bias treatment. The 
    treatment was conducted with a probe station and a hot plate set at 353 K, following the process described in Refs. \cite{pappas2024alternating,wang2024precision}, but 
    with +/- 10 V bias pulses and a one-second dwell time between pulses for a total duration of 30 hours. After treatment, the device was mounted in a dilution 
    refrigerator, and the transmission spectrum was measured at a base temperature of 10 mK.
    The extracted dipole moments ranged from 0.15 to 0.65 e$\rm{\AA}$, with a dipole density of 106 $\pm$ 27 $\rm{TLS}/(\mu m^3\rm{GHz})$. Data processing for this figure is described in Supplemental Material \ref{SupSec:TLS_fitting_details}.}
    \label{Supfig:After_RT_cycle_and_ABAA}
\end{figure*}

Supplemental Figure \ref{Supfig:After_RT_cycle_and_ABAA} presents the results obtained after the device underwent thermal cycling to 
300 K (room temperature) and was subjected to an alternating bias treatment, as described in Methods and detailed in \cite{pappas2024alternating}. 
During the experiment, the device was mounted on a voltage probe station, where an alternating bias treatment, as described in Methods and 
detailed in \cite{pappas2024alternating}, was applied at 353 K for 30 hours. After cooling the device back down to 10 mK,  
TLSs with dipole moments of 0.15 - 0.65 e$\rm{\AA}$ and a density of 106 $\pm$ 27 $\rm{TLS}/(\mu m^3\rm{GHz})$ were measured.

These results indicate that the TLS energy landscape remained consistent with measurements taken after the 10 K thermal cycle 
(see Table \ref{tabdipoledens} and Fig. \ref{fig:Pre_post_ABAA_histogram} in the main paper). Specifically, the alternating 
bias treatment did not significantly alter TLS properties, as evidenced 
by the unchanged loss tangent value ($\tan\delta^0_{\mathrm{TLS}}$ = 1.64$\times$10$^{-3}$). This suggests that the alternating bias treatment preserves the TLS 
landscape under the tested conditions and does not significantly affect the strongly coupled TLSs.

\end{document}